\documentclass[pre,twocolumn,superscriptaddress]{revtex4-1}  %floatfix

\usepackage[paperwidth=210mm,paperheight=297mm,centering,hmargin=2cm,vmargin=2.5cm]{geometry}

\usepackage[pdftex]{graphicx}
\usepackage{amsmath,amsfonts,amssymb}
\usepackage{morefloats}
\usepackage{slashbox}
\usepackage{natbib}
\usepackage{array}
\usepackage{url}
\usepackage{dsfont}

\usepackage{hyperref}
\usepackage{xcolor} 

\definecolor{bluemoi}{rgb}{0.25,0.50 ,0.75} 

\hypersetup{
  bookmarksopen=true,
  pdftitle=" ",
  pdfauthor=" ", 
  pdfsubject=" ", 
  pdftoolbar=true, % toolbar hidden
  pdfmenubar=true, %menubar shown
  pdfhighlight=/O, %effect of clicking on a link
  colorlinks=true, %couleurs sur les liens hypertextes
  pdfpagemode=UseNone, %aucun mode de page
  pdfpagelayout=SinglePage, %ouverture en simple page
  pdffitwindow=true, %pages ouvertes entierement dans toute la fenetre
  linkcolor=bluemoi, %couleur des liens hypertextes internes
  citecolor=bluemoi, %couleur des liens pour les citations
  urlcolor=bluemoi %couleur des liens pour les url
}

\makeatletter
\renewcommand{\figurename}{\sf \textbf{Figure}}
\renewcommand{\thefigure}{\arabic{figure}}
\renewcommand{\fnum@figure}{\sf\textbf{\figurename}~\textbf{\thefigure}}
\renewcommand{\tablename}{\sf\textbf{Table}}
\renewcommand{\thetable}{\arabic{table}}
\renewcommand{\fnum@table}{\sf\textbf{\tablename}~\textbf{\thetable}}
\makeatother

\begin{document}

\title{Systematic comparison of trip distribution laws and models}

\author{Maxime Lenormand}\affiliation{Instituto de F\'isica Interdisciplinar y Sistemas Complejos IFISC (CSIC-UIB), Campus UIB, 07122 Palma de Mallorca, Spain}
\author{Aleix Bassolas}\affiliation{Instituto de F\'isica Interdisciplinar y Sistemas Complejos IFISC (CSIC-UIB), Campus UIB, 07122 Palma de Mallorca, Spain}
\author{Jos\'e J. Ramasco}\affiliation{Instituto de F\'isica Interdisciplinar y Sistemas Complejos IFISC (CSIC-UIB), Campus UIB, 07122 Palma de Mallorca, Spain}

\begin{abstract} 
Trip distribution laws are basic for the travel demand characterization needed in transport and urban planning. Several approaches have been considered in the last years. One of them is the so-called gravity law, in which the number of trips is assumed to be related to the population at origin and destination and to decrease with the distance. The mathematical expression of this law resembles Newton's law of gravity, which explains its name.  Another popular approach is inspired by the theory of intervening opportunities which argues that the distance has no effect on the destination choice, playing only the role of a surrogate for the number of intervening opportunities between them. In this paper, we perform a thorough comparison between these two approaches in their ability at estimating commuting flows by testing them against empirical trip data at different scales and coming from different countries. Different versions of the gravity and the intervening opportunities laws, including the recently proposed radiation law, are used to estimate the probability that an individual has to commute from one unit to another, called trip distribution law. Based on these probability distribution laws, the commuting networks are simulated with different trip distribution models. We show that the gravity law performs better than the intervening opportunities laws to estimate the commuting flows, to preserve the structure of the network and to fit the commuting distance distribution although it fails at predicting commuting flows at large distances. Finally, we show that the different approaches can be used in the absence of detailed data for calibration since their only parameter depends only on the scale of the geographic unit. 
\end{abstract}

\maketitle

%\section*{INTRODUCTION}

Everyday, billions of individuals around the world travel. These movements form a socio-economic complex network, backbone for the transport of people, goods, money, information or even diseases at different spatial scales. The study of such spatial networks is consequently the subject of an intensive scientific activity \cite{Barthelemy2011}. Some examples include the estimation of population flows \cite{Murat2010,Gargiulo2012,Simini2012,Lenormand2012,Thomas2013,Lenormand2014,Yang2014,Sagarra2015}, transport planning and modeling \cite{Rouwendal2004,Ortuzar2011}, spatial network analysis \cite{DeMontis2007,DeMontis2010}, study of urban traffic \cite{DeMontis2007} and modeling of the spreading of infectious diseases \cite{Viboud2006,Balcan2009,Tizzoni2014}.

Trip distribution modeling is thus crucial for the prediction of population movements, but also for an explanatory purpose, in order to better understand the mechanisms of human mobility. There are two major approaches for the estimation of trip distribution at an aggregate level. The traditional gravity approach, in analogy with the Newton's law of gravitation, is based on the assumption that the amount of trips between two locations is related to their populations and decays with a function of the distance \cite{Carey1858, Zipf1946,Wilson1970, Erlander1990}. In contrast to the gravity law, the Stouffer's law of intervening opportunities \cite{Stouffer1940} hinges on the assumption that the number of opportunities plays a more important role in the location choices than the distance, particularly in the case of migration choices. The original law proposed by Stouffer has been reformulated by Schneider \cite{Schneider1959} and extensively studied since then \cite{Heanue1966,Ruiter1967,Wilson1970,Haynes1973,Fik1990,Akwawua2001}. The two approaches have been widely compared during the second half of the twentieth century \cite{David1961,Pyers1966,Lawson1967,Zhao2001} showing that generally both approaches performed comparably. However, the simplicity of the mathematical form of the gravity approach appears to have weighted in its favor \cite{Ortuzar2011}. Indeed, the gravity approach has been extensively used in the past few decades to model, for instance, flows of population \cite{Viboud2006,Griffith2009,Balcan2009,Murat2010,Gargiulo2012,Lenormand2012,Thomas2013,Masucci2013,Liang2013,Lenormand2014,Tizzoni2014,Liu2014}, spatial accessibility to health services \cite{Luo2003}, volume of international trade \cite{Anderson1979,Bergstrand1985}, traffic in transport networks \cite{Jung2008,Kaluza2010} and phone communications \cite{Krings2009}.  

\begin{figure*}
\begin{center}
\includegraphics[width=14cm]{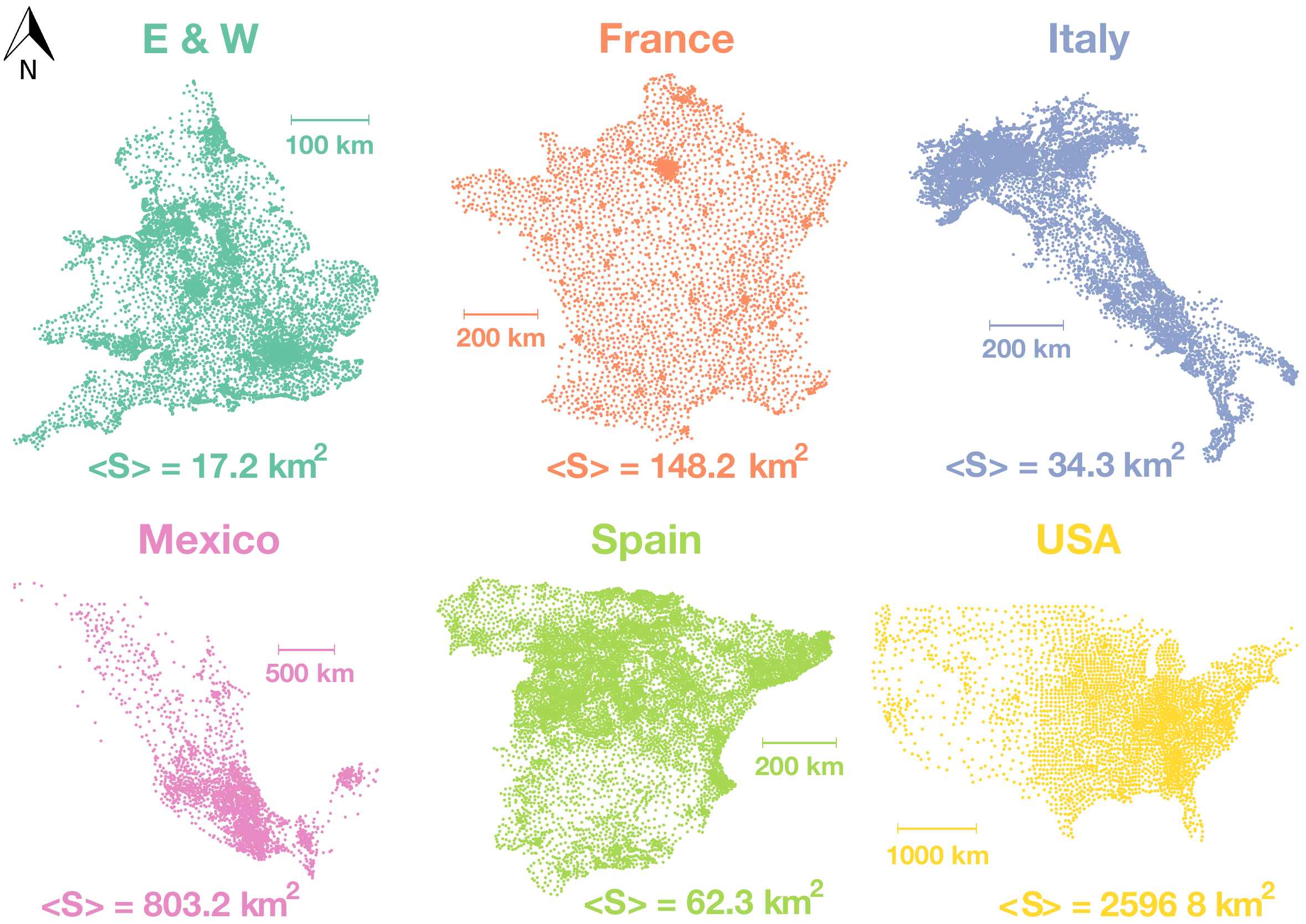}
\caption{\textbf{Position of the units' centroids for the six countries.} $\langle S \rangle$ represents the average surface of the census units (i.e. municipalities, counties or wards). \label{Fig1}}
\end{center}	
\end{figure*}

However, the concept of intervening opportunities has recently regained in popularity thanks to the recently proposed radiation approach \cite{Simini2012,Simini2013,Ren2014,Yang2014}. This approach is inspired by a simple diffusion model where the amount of trips between two locations depends on their populations and the number of opportunities between them. The gravity law and the radiation law have been compared several times during the last years giving the superiority to either of the approaches depending on the study \cite{Simini2012,Lenormand2012,Masucci2013,Liang2013,Yang2014}. Two main issues can be identified in these comparisons. First, the inputs used to simulate the flows are not always identical. For example, in the comparison proposed in \cite{Masucci2013}, the gravity law tested takes as input the population, whereas the radiation law is based on the number of jobs. Second, in all these studies, the models used to generate the trips from the radiation and the gravity laws are not constrained in the same way. The radiation models are always production constrained, this means that the number of trips, or at least an estimation of the number of trips generated by census unit, is preserved. The models used to generate the trips with the gravity laws can be either, unconstrained \cite{Simini2012,Masucci2013}, only the total number of trips is preserved or doubly constrained \cite{Lenormand2012,Yang2014}, both the trips produced and attracted by a census unit are preserved. Therefore, to fairly compare different approaches the same input data must be used and, most importantly, we need to differentiate the law, gravity or intervening opportunities, and the modeling framework used to generate the trips from this law. Indeed, both the gravity laws and the intervening opportunities laws can be expressed as a probability to move from one place to another, called trip distribution law, and based on these probability distributions, the total number of trips can then be simulated using different trip distribution models including different level of constraints. 

In this work, we test and compare, in a systematic and rigorous way, gravity and intervening opportunities laws against commuting census data coming from six different countries using four different constrained models to generate the networks: unconstrained model, single constrained models (production or attraction) and the well-known doubly constrained model. For the gravity law, since the form of the distance decay functions may vary from one study to another \cite{Fotheringham1981,Viboud2006,Vries2009,Balcan2009,Barthelemy2011,Lenormand2014,Chen2015} both the power and the exponential forms are tested to model the impact of the distance. The intervening opportunities law is given by the Schneider's version of the Stouffer's original law as it is usually the case. We also considered two versions of the radiation law, the original free-parameter model \cite{Simini2012} and the extended version proposed in \cite{Yang2014}. The simulated networks are compared with the observed ones on different aspects showing that, globally, the gravity law with an exponential distance decay function outperforms the other laws in the estimation of commuting flows, the conservation of the commuting network structure and the fit of the commuting distance distribution even if it fails at predicting commuting flows at large distances. Finally, we show that the different laws can be used in absence of detailed data for calibration since their only parameter depends only on the scale of the geographic census unit.

\section*{DATA}

In this study, the trip distribution laws and models are tested against census commuting data of six countries: England and Wales, France, Italy, Mexico, Spain and the United States of America (hereafter called E\&W, FRA, ITA, MEX, SPA and USA, respectively) and two cities: London and Paris (hereafter called LON and PAR, respectively).  

\begin{itemize}
	\item The England \& Wales dataset comes from the $2001$ Census in England and Wales made available by the Office for National Statistics (data available online at \url{https://www.nomisweb.co.uk/query/construct/summary.asp?mode=construct&version=0&dataset=124}).
	\item The French dataset was measured for the $1999$ French Census by the French Statistical Institute (data available upon request at \url{http://www.cmh.ens.fr/greco/adisp_eng.php}).
	\item The Italian's commuting network was extracted from the $2001$ Italian Census by the National Institute for Statistics (data available upon request at \url{http://www.istat.it/it/archivio/139381}).
	\item Data on commuting trips between Mexican's municipalities in $2011$ are based on a microdata sample coming from the Mexican National Institute for Statistics (data available online at \url{http://www3.inegi.org.mx/sistemas/microdatos/default2010.aspx}).
	\item The Spanish dataset comes from the $2001$ Spanish Census made available by the Spanish National Statistics Institute (data available upon request at \url{http://www.ine.es/en/censo2001/index_en.html}).
	\item Data on commuting trips between United States counties in $2000$ comes from the United State Census Bureau (data available online at \url{https://www.census.gov/population/www/cen2000/commuting/index.html}).
\end{itemize}

Each case study is divided into $n$ census units of different spatial scale: from the Output Area in London with an average surface of $1.68\mbox{ km}^2$ to the counties in the United States with an average surface of $2596.8\mbox{ km}^2$. See Table \ref{tab1} for a detailed description of the datasets.

\begin{table*}[!ht]
	\caption{Presentation of the datasets \label{tab1}} 
	\label{Datasets}
			\begin{tabular}{>{}m{3cm}>{\centering}m{4cm}>{\centering}m{3cm}m{3cm}<{\centering}}
			  \hline
				\textbf{Case study} & \textbf{Number of units} & \textbf{Number of links} & \textbf{Number of Commuters}\\
				\hline
				England \& Wales & 8,846 wards & 1,269,396 & 18,374,407\\
				France & 3,645 cantons & 462,838 & 12,193,058\\
				Italy & 7,319 municipalities & 419,556 & 8,973,671\\
				Mexico & 2,456 municipalities & 60,049 & 603,688 \\
				Spain & 7,950 municipalities & 261,084 & 5,102,359\\
				United State & 3,108 counties & 161,522 & 34,097,929\\
				London & 4,664 Output Areas & 750,943 & 4,373,442\\
				Paris & 3,185 municipalities & 277,252 & 3,789,487\\
				\hline
	  	\end{tabular}
\end{table*}

\begin{figure}[!ht]
\begin{center}
\includegraphics[width=\linewidth]{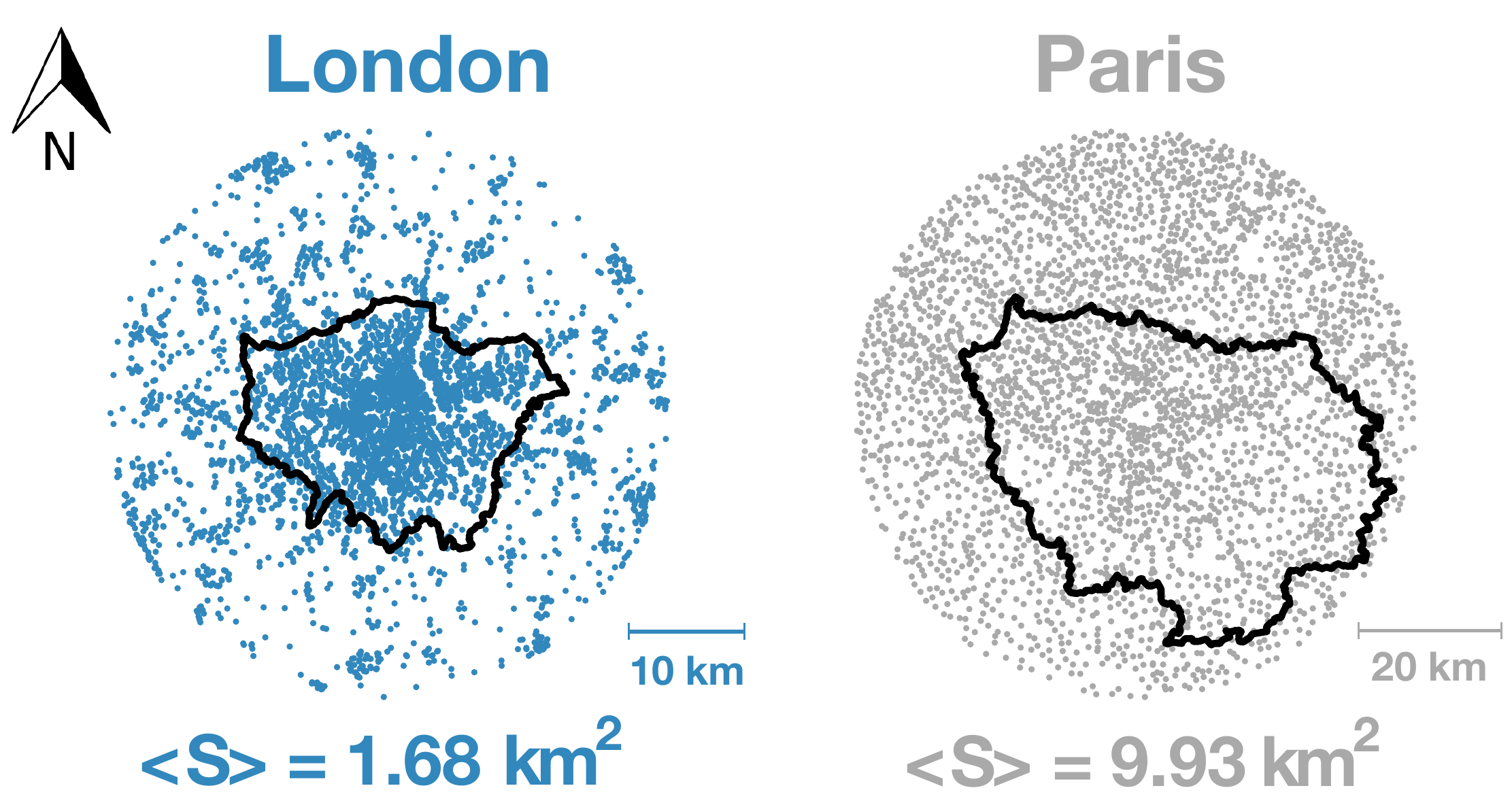}
\caption{\textbf{Position of the units' centroids around London (left) and Paris (right).}  The black contours represent the boundaries of the Greater London Authority (left) and the french \textit{d{\'e}partement} {I}le de France (right). $\langle S \rangle$ represents the average unit surface. \label{Fig2}}
\end{center}
\end{figure}

Figures \ref{Fig1} and \ref{Fig2} display the centroids of the census units for the eight case studies. For each unit, the statistical offices provide the following information:

\begin{itemize}
	\item $T_{ij}$, the number of trips between the census units $i$ and $j$ (i.e. number of individuals living in $i$ and working in $j$);
	\item $d_{ij}$, the great-circle distance between the unit $i$ and the unit $j$ computed with the Haversine formula;
	\item $m_i$, the number of inhabitants in unit $i$.
\end{itemize}

In this work we consider only inter-unit flows (i.e. $T_{ii}=0$), mainly because it is not possible to estimate intra-units flows with the radiation laws \footnote{~Note that it is possible to estimate intra-unit flows with the gravity laws by approximating intra-unit distances with, for example, half the square root of the unit's area or half the average distance to the nearest neighbors.}. We note $N=\sum_{i,j=1}^n T_{ij}$ the total number of commuters, $O_i=\sum_{j=1}^n T_{ij}$ the number of out-commuters (i.e. number of individuals living in $i$ and working in another census unit) and $D_j=\sum_{i=1}^n T_{ij}$ the number of in-commuters (i.e. number of individuals working in $j$ and living in another census unit ).

\section*{COMPARISON OF TRIP DISTRIBUTION LAWS AND MODELS}

The purpose of the trip distribution models is to split the total number of trips $N$ in order to generate a trip table $\tilde{T}=(\tilde{T}_{ij})_{1 \leq i,j \leq n}$ of the estimated number of trips form each census area to every other. Note that by trip we are referring to commuting travels from home to work, there is a return trip not considered in $\tilde{T}$ and $N$ is also equivalent to the number of unique commuters. The trip distribution depends on, on one hand, the characteristics of the census units and the way they are spatially distributed, and, on the other hand, the level of constraints required by the model. Therefore, to fairly compare different trip distribution modeling approaches we have to consider separately the law used to calculate the probability to observe a trip between two census units, called trip distribution law, and the trip distribution model used to generate the trip allocation from this law.   

\subsection*{Gravity and intervening opportunities laws}

The purpose of this study is to test the capacity of both the gravity and the intervening opportunities approaches to estimate the probability $p_{ij}$ that out of all the possible travels in the system we have one between the census unit $i$ and $j$. This probability is asymmetric in $i$ and $j$ as the flows themselves, and, by convention, the self-loops are excluded of the analysis $p_{ii}=0$. This probability is normalized to all possible couples of origins and destinations, $\sum_{i,j=1}^n p_{ij} =1$. Note that $p_{ij}$ does not refer to the conditional probability of a trip starting in $i$ finishes in $j$ $\mathbb{P}(1|i,j)$. There exists a relation between both of them:

\begin{equation}
p_{ij} = \mathbb{P}(i) \, \mathbb{P}(1|i,j) 
\end{equation}

where $\mathbb{P}(i)$ stands for the probability of a trip starting in $i$. $\mathbb{P}(1|i,j)$ will appear later for the intervening opportunities laws as a function of the populations of origin $m_i$, destination $m_j$ and the number of opportunities between them $s_{ij}$, $\mathbb{P}(1|m_i,m_j,s_{ij})$, but the basis of our analysis will be $p_{ij}$.

\subsubsection*{Gravity laws}

In the simplest form of the gravity approach, the probability of commuting between two units $i$ and $j$ is proportional to the product of the origin population $m_i$ and destination population $m_j$, and inversely proportional to the travel cost between the two units:

\begin{equation}
p_{ij} \propto m_i \, m_j \, f(d_{ij}),\,\,\,\,\,\,i\ne j  
\label{grav}
\end{equation}

The travel cost between $i$ and $j$ is usually modeled with an exponential distance decay function, 

\begin{equation}
 f(d_{ij})=e^{-\beta \, d_{ij}} 
 \label{exp}
\end{equation}

or a power distance decay function,

\begin{equation}
   f(d_{ij})={d_{ij}}^{-\beta}  \label{pow} 
\end{equation}

As mentioned in \cite{Barthelemy2011}, the form of the distance decay function can change according to the dataset, therefore, both the exponential and the power forms are considered in this study. In both cases, the importance of the distance in commuting choices is adjusted with a parameter $\beta$ with observed data.

\subsubsection*{Intervening opportunities laws}

In the intervening opportunity approach, the probability of commuting between two units $i$ and $j$ is proportional to the origin population $m_i$ and to the conditional probability that a commuter living in unit $i$ with population $m_i$ is attracted to unit $j$ with population $m_j$, given that there are $s_{ij}$ job opportunities in between. The conditional probability $\mathbb{P}(1|m_i,m_j,s_{ij})$ needs to be normalized to ensure that all the trips end in the region of interest.

\begin{equation}
   p_{ij} \propto m_i \, \frac{\mathbb{P}(1|m_i,m_j,s_{ij})}{\sum_{k=1}^n\mathbb{P}(1|m_i,m_k,s_{ik})},\,\,\,\,\,\,i\ne j  \label{IO}
\end{equation}

In the Schneider's version of the intervening opportunities approach the conditional probability is given by

\begin{equation}
  \mathbb{P}(1|m_i,m_j,s_{ij})=  e^{\displaystyle -\gamma s_{ij}}-e^{\displaystyle -\gamma (s_{ij}+m_j)}
  \label{schneider}
\end{equation}

where $s_{ij}$ is the number of opportunities (approximated by the population in this case) in a circle of radius $d_{ij}$ centered in $i$ (excluding the source and destination). The parameter $\gamma$ can be seen as a constant probability of accepting an opportunity destination. Note that in this version the number of opportunities $m_i$ at the origin is not taken into account.

More recently, \cite{Simini2012} reformulated the Stouffer's intervening opportunities law in terms of radiation and absorption processes. This model is inspired by a diffusion model where each individual living in an unit $i$ has a certain probability of being ''absorbed'' by another unit $j$ according to the spatial distribution of opportunities. The original radiation model is free of parameters and, therefore, it does not require calibration. The conditional probability  $\mathbb{P}(1|m_i,m_j,s_{ij})$ is expressed as:

\begin{equation}
   \mathbb{P}(1|m_i,m_j,s_{ij})=\frac{m_i \, m_j}{(m_i+s_{ij})\, (m_i+m_j+s_{ij})} \label{rad}
\end{equation}

This conditional probability needs to be normalized because the probability for an individual living in a census unit $i$ of being absorbed by another census unit is not equal to $1$ in case of finite system but equal to $1-\frac{m_i}{M}$ where $M$ is the total population \cite{Masucci2013}. Some recent works have shown that the model fails to describe human mobility compared to more classic approaches particularly on a small scale \cite{Lenormand2012,Masucci2013, Liang2013}. To circumvent these limitations, an extended radiation model has been proposed by \cite{Yang2014}. In this extended version, the probability $\mathbb{P}(1|m_i,m_j,s_{ij})$ is derived under the survival analysis framework introducing a parameter $\alpha$ to control the effect of the number of job opportunities between the source and the destination on the job selection,
 
\begin{equation}
   \hspace*{-0.6cm} \mathbb{P}(1|m_i,m_j,s_{ij})=\frac{[{(m_i+m_j+s_{ij})}^\alpha - {(m_i+s_{ij})}^\alpha]\, ({m_i}^\alpha + 1)}{[{(m_i+s_{ij})}^\alpha + 1]\, [{(m_i+m_j+s_{ij})}^\alpha + 1]}  \label{extrad}
\end{equation}

\subsection*{Constrained models}

After the description of the probabilistic laws, the next step is to materialize the people commuting. 
The purpose is to generate the commuting network $\tilde{T}=(\tilde{T}_{ij})_{1 \leq i,j \leq n}$ by drawing at random $N$ trips from the trip distribution law $(p_{ij})_{1 \leq i,j \leq n}$ respecting different level of constraints according to the model. We are going to consider four different types of models:

\begin{enumerate}

\item {\it Unconstrained model.} The only constraint of this model is to ensure that the total number of trips $\tilde{N}$ generated by the model is equal to the total number of trips $N$ observed in the data. In this model, the $N$ trips are randomly sample from the multinomial distribution,

\begin{equation}
	\displaystyle \mathcal{M}\left(N,\left(p_{ij}\right)_{1 \leq i,j \leq n}\right)  \label{NC}
\end{equation}

\item {\it Production constrained model.} This model ensures that the number of trips ''produced'' by a census unit is preserved. For each unit $i$, $O_i$ trips are produced from the multinomial distribution,

\begin{equation}
	\displaystyle \mathcal{M}\left(O_i,\left(\frac{p_{ij}}{\sum_{k=1}^n p_{ik}}\right)_{1 \leq j \leq n}\right)  \label{PCM}
\end{equation}

\item {\it Attraction constrained model.} This model ensures that the number of trips ''attracted'' by a unit is preserved. For each census unit $j$, $D_j$ trips are attracted from the multinomial distribution,

\begin{equation}
	\displaystyle \mathcal{M}\left(D_j,\left(\frac{p_{ij}}{\sum_{k=1}^n p_{kj}}\right)_{1 \leq i \leq n}\right)  \label{ACM}
\end{equation}

\item {\it Doubly constrained model.} This model, also called production-attraction constrained model ensures that both the trips attracted and generated by a census unit are preserved using two balancing factors $K_i$ and $K_j$ calibrated with the \textit{Iterative Proportional Fitting} procedure \cite{Deming1940}. The relation between $K_i$, $K_j$, $p_{ij}$ and the trip flows is given by

\begin{equation}
  \left\{ 
    \begin{array}{l}  
		    \tilde{T}_{ij} = K_i \, K_j \, p_{ij} \\ 
        \sum_{j=1}^n \tilde{T}_{ij}=O_i, \,\,\sum_{i=1}^n \tilde{T}_{ij}=D_j  \\											 
    \end{array} 
	\right.
   %\displaystyle \tilde{T}_{ij} = K_i \, K_j \, p_{ij} 
	 \label{DC}
\end{equation}

Unlike the unconstrained and single constrained models, the doubly constrained model is a deterministic model. Therefore, the simulated network $\tilde{T}$ is a fully connected network in which the flows are real numbers instead of integers. This can be problematic since we want to study the capacity of both the gravity and the radiation approaches to preserve the topological structure of the original network. To bypass this limitation $N$ trips are randomly sample from the multinomial distribution,

\begin{equation}
	\displaystyle \mathcal{M}\left(N,\left(\frac{\tilde{T}_{ij}}{\sum_{k,l=1}^n \tilde{T}_{kl}}\right)_{1 \leq i,j \leq n}\right)  \label{DC2}
\end{equation}

\end{enumerate}

\subsection*{Goodness-of-fit measures}

\paragraph*{Common part of commuters} We calibrate the parameters $\beta$, $\gamma$ and $\alpha$ using the common part of commuters (CPC) introduced in \cite{Gargiulo2012,Lenormand2012}:

\begin{equation}
   \displaystyle CPC(T,\tilde{T}) = \frac{2\sum_{i,j=1}^n min(T_{ij},\tilde{T}_{ij})}{\sum_{i,j=1}^n T_{ij} + \sum_{i,j=1}^n \tilde{T}_{ij}}  \label{CPC}
\end{equation}

This indicator is based on the S{\o}rensen index \cite{Sorensen1948}. It varies from $0$, when no agreement is found, to $1$, when the two networks are identical.  In our case, the total number of commuters $N$ is preserved, therefore the Equation (\ref{CPC}) can be simplified to

\begin{equation}
   \displaystyle CPC(T,\tilde{T}) = 1 - \frac{1}{2}\frac{\sum_{i,j=1}^n |T_{ij}-\tilde{T}_{ij}|}{N}  \label{CPC2}
\end{equation}

which represents the percentage of good prediction as defined in \cite{Lenormand2013}.  

In order to assess the robustness of the results regarding the choice of goodness-of-fit measures, we also test the results obtained with the normalized root mean square error,

\begin{equation}
   \displaystyle NRMSE(T,\tilde{T}) = \frac{\sum_{i,j=1}^n (T_{ij}-\tilde{T}_{ij})^2}{\sum_{i,j=1}^n T_{ij}}  \label{NRMSE}
\end{equation}

and the information gain statistic, 

\begin{equation}
   \displaystyle I(T,\tilde{T}) = \sum_{i,j=1}^n \frac{T_{ij}}{N}ln\left(\frac{T_{ij}}{\tilde{T}_{ij}}\right)  \label{I}
\end{equation}

\paragraph*{Common part of links} The ability of the models to recover the topological structure of the original network can be assessed with the common part of links (CPL) defined as

\begin{equation}
   \displaystyle CPL(T,\tilde{T}) = \frac{2\sum_{i,j=1}^n \mathds{1}_{T_{ij}>0} \cdot \mathds{1}_{\tilde{T}_{ij}>0}}{\sum_{i,j=1}^n \mathds{1}_{T_{ij}>0} + \sum_{i,j=1}^n \mathds{1}_{\tilde{T}_{ij}>0}} \label{CPL}
\end{equation}

where $\mathds{1}_X$ is equal to one if the condition $X$ is fulfilled and zero otherwise. The common part of links measures the proportion of links in common between the simulated and the observed networks (i.e. links such as $T_{ij}>0$ and $\tilde{T}_{ij}>0$). It is null if there is no link in common and one if both networks are topologically equivalent. 

\begin{figure*}
 \begin{center}
\includegraphics[width=12cm]{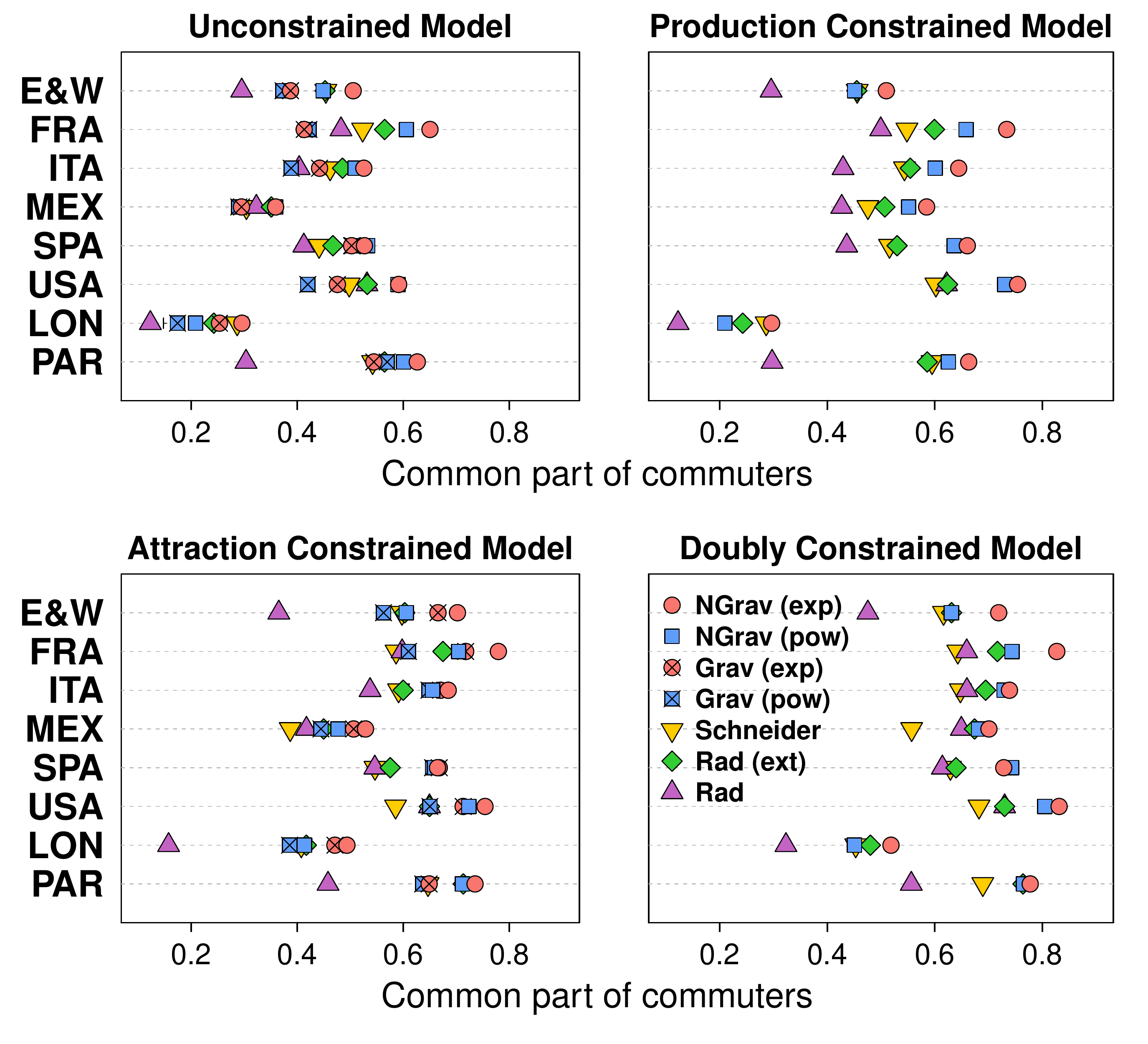}
	\caption{\textbf{Common part of commuters according to the unconstrained models, the gravity and intervening opportunities laws for the eight case studies.} The circles represent the normalized gravity law with the exponential distance decay function (the circles with a cross inside represent the original version); The squares represent the normalized gravity law with the power distance decay function (the squares with a cross inside represent the original version); The point down triangles represent the Schneider's intervening opportunities law; The green diamonds represent the extended radiation law; The purple triangles represent the original radiation law. Error bars represent the minimum and the maximum values observed in the $100$ realizations but in most cases they are too close to the average to be seen. \label{Fig3}}
\end{center}
\end{figure*}

\paragraph*{Common part of commuters according to the distance} In order to measure the similarity between the observed commuting distance distribution and the ones simulated with the models, we introduce the common part of commuters according to the distance (CPC$_d$). Let us consider $N_k$ the number of individuals having a commuting distance in the bin between $2k-2$ and $2k$ kms. The CPC$_d$ is equal to the CPC based on $N_k$ instead of $T_{ij}$ 

\begin{equation}
   \displaystyle CPC_d(T,\tilde{T}) = \frac{\sum_{k=1}^{\infty} min(N_{k},\tilde{N}_{k})}{N}  \label{CPCd}
\end{equation}

\section*{RESULTS}

In this section, we compare the five laws: gravity with an exponential or a power distance decay function, the Schneider's intervening opportunities law and the original and the extended radiation laws. We test these laws against empirical data coming from eight different case studies using four constrained models to estimate the flows. For each constrained model, the parameters $\beta$, $\gamma$ and $\alpha$ are calibrated so as to maximize the CPC. Since the models are stochastic, we consider an average CPC value measured over $100$ replications of the trip distribution. Similarly, all the goodness-of-fit measures are obtained by calculating the average measured over $100$ network replications. It is important to note that the networks generated with the constrained models are very stable, the stochasticity of the models does not affect the statistical properties of the network. Therefore, the goodness-of-fit measures does not vary much with the different realizations of the multinomial sampling. For example, within the $100$ network instances for all models and case studies, the CPC varies, at most, by $0.09\%$ around the average.

\begin{figure*}
\begin{center}
\includegraphics[width=12cm]{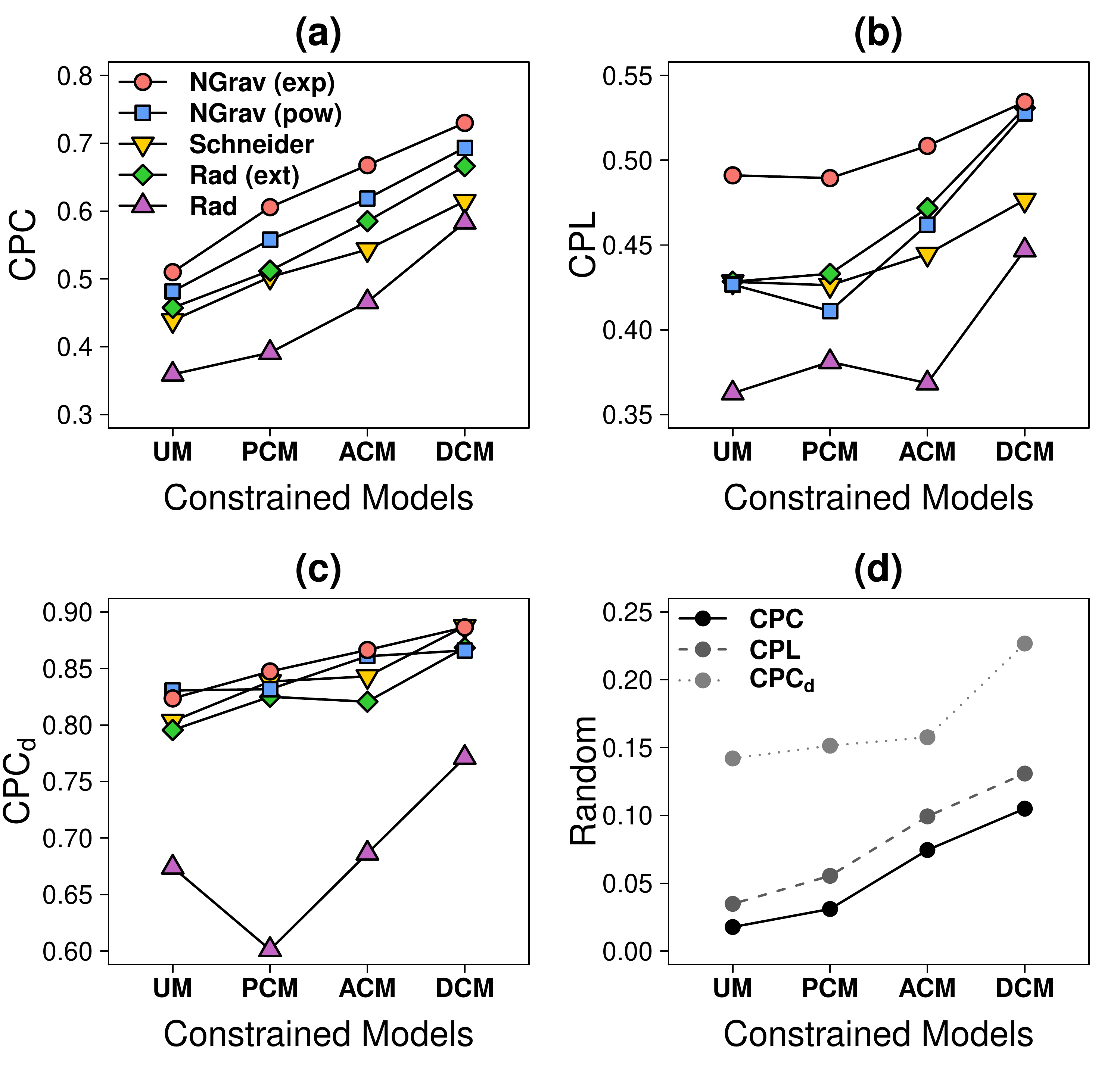}
\caption{\textbf{Performance of the unconstrained model (UM), the production constrained model (PCM), the attraction constrained model (ACM) and the doubly constrained model (DCM) according to the gravity and the intervening opportunities laws (a)-(c) and a uniform distribution (d).} (a) Average CPC. (b) Average CPL. (c) Average CPC$_d$. The red circles represent the normalized gravity law with the exponential distance decay function; The blue squares represent the normalized gravity law with the power distance decay function; The point down triangles represent the Schneider's intervening opportunities law; The green diamonds represent the extended radiation law; The purple triangles represent the original radiation law. The grey point down triangles represent the uniform distribution, form dark to light grey, the CPC, the CPL and the CPC$_d$. \label{Fig4}}
\end{center}
\end{figure*}

\subsection*{Estimation of commuting flows}

Figure \ref{Fig3} displays the common part of commuters obtained with the different laws and models for the eight case studies. Globally, the gravity laws give better results than the intervening opportunities laws. For the gravity laws, the results improve with the exponential rather than with the power distance decay function. For the intervening opportunities laws, the extended radiation law outperforms the original one and achieves slightly better results than the Schneider law. In the top left panel, we observe the results for the unconstrained model. In this case, the extended radiation law and the Schneider law give better results than the gravity ones for most case studies. However, these better performances are due to the normalization factor used in Equation \ref{IO}. Indeed, this normalization implies that the probability of having a trip originating in a census unit $i$ is proportional to the population of $i$, which is not necessarily the case for the gravity laws. If we use the same type of normalization for the gravity trip distribution law $p_{ij}$ (Equation \ref{NGrav}), we observe that the ''normalized'' gravity laws give better results than the intervening opportunities laws. In the following, we will refer to the normalized version when mentioning the gravity law.

\begin{equation}
   p_{ij} \propto m_i \frac{m_j f(d_{ij})}{\sum_{k=1}^n m_k f(d_{ik})} ,\,\,\,\,\,\,i\ne j 
	 \label{NGrav}
\end{equation}

To compare the constrained models performances, we plot in Figure \ref{Fig4}a the CPC obtained with the four models according to the laws averaged over the eight case studies. As expected, more constrained the model is, higher the CPC becomes. Unconstrained models are able to reproduce on average around $45\%$ of the observed commuting network against $65\%$ for the doubly constrained model. It is interesting to note that, the attraction constrained model gives better results than the production constrained model. This can be explained by the fact that the job demand is easier to estimate than the job offer, which can be related to extra economic questions. This is in agreement with the results obtained with a uniform distribution ($p_{ij}\propto 1$) plotted in Figure \ref{Fig3}d.  

Although the results obtained with the normalized root mean square error and the information gain statistic are very similar to the ones obtained with the CPC, it is worth noting that globally the extended radiation law gives smaller normalized root mean square error values than the normalized gravity laws with the unconstrained model (see Table \ref{tab2} for more details about the laws exhibiting the best performances). 

\begin{figure*}
\begin{center}
\includegraphics[width=12cm]{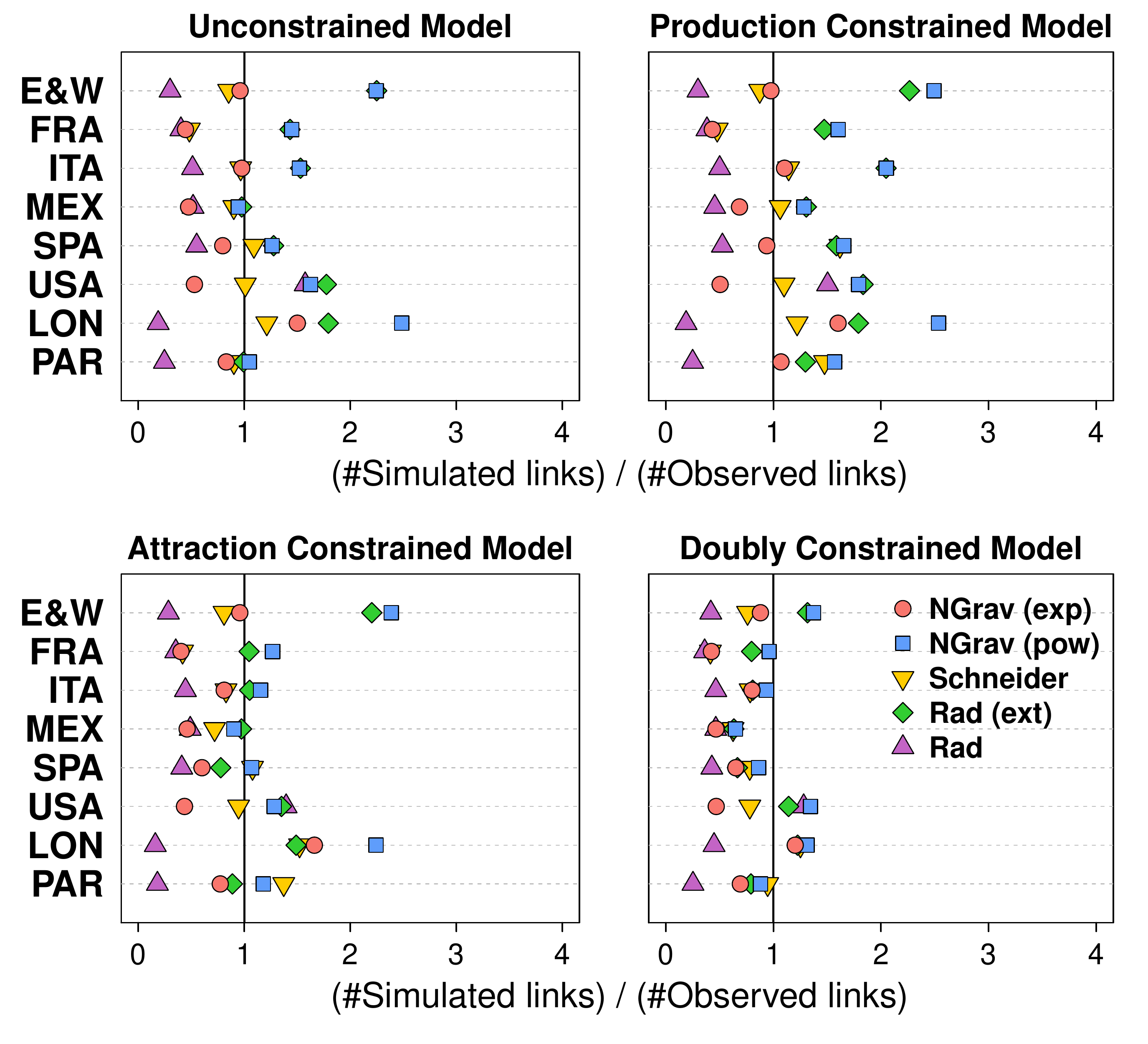}
\caption{\textbf{Ratio between the simulated and the observed number of links according to the unconstrained models, the gravity and intervening opportunities laws for the eight case studies.} The red circles represent the normalized gravity law with the exponential distance decay function; The blue squares represent the normalized gravity law with the power distance decay function; The point down triangles represent the Schneider's intervening opportunities law; The green diamonds represent the extended radiation law; The purple triangles represent the original radiation law. Error bars represent the minimum and the maximum but in most cases they are too close to the average to be seen. \label{Fig5}}
\end{center}
\end{figure*} 

\begin{figure*}
 \begin{center}
\includegraphics[width=14cm]{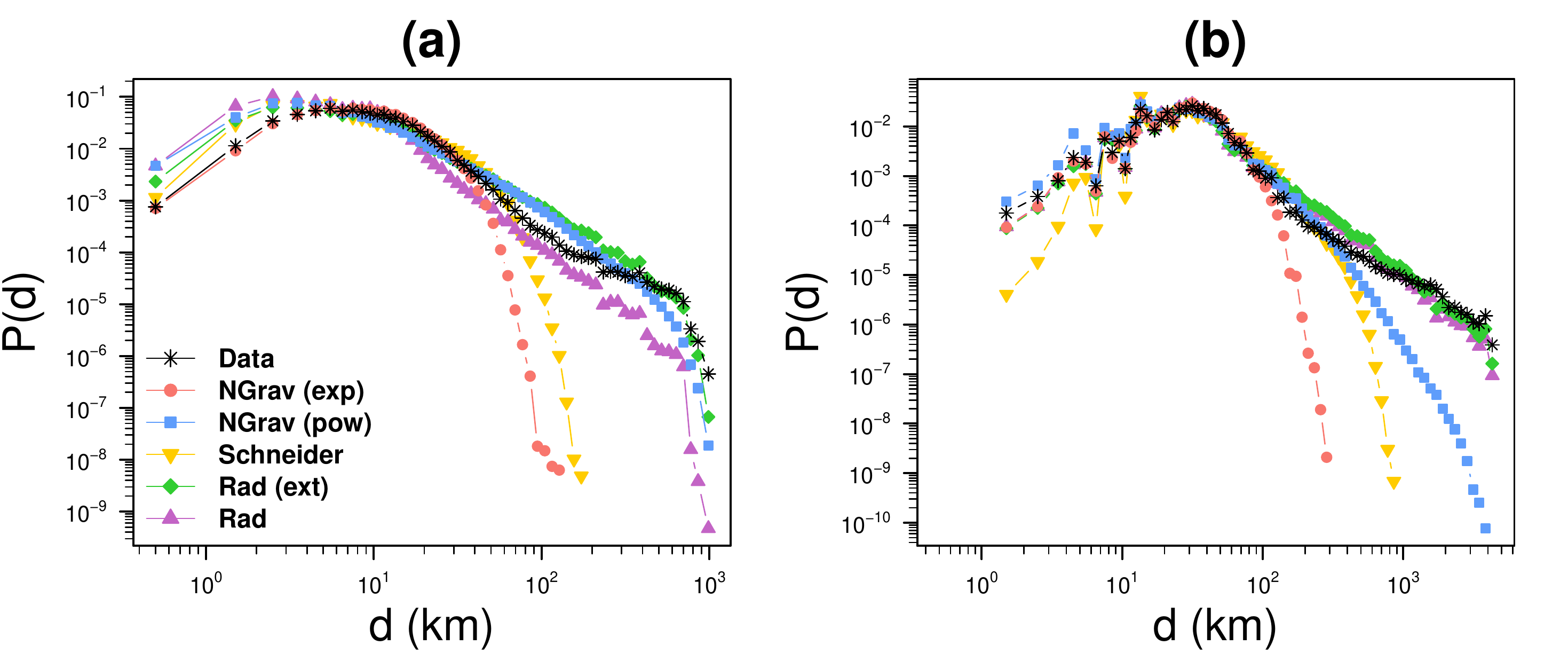}
\caption{\textbf{Probability density function of the commuting distance distribution observed in the data and simulated with the production constrained model.} (a) France and (b) United States. The red circles represent the normalized gravity law with the exponential distance decay function; The blue squares represent the normalized gravity law with the power distance decay function; The point down triangles represent the Schneider's intervening opportunities law; The green diamonds represent the extended radiation law; The purple triangles represent the original radiation law. The black stars represent the census data. \label{Fig6}}
\end{center}
\end{figure*}

\begin{figure*}
\begin{center}
\includegraphics[width=14cm]{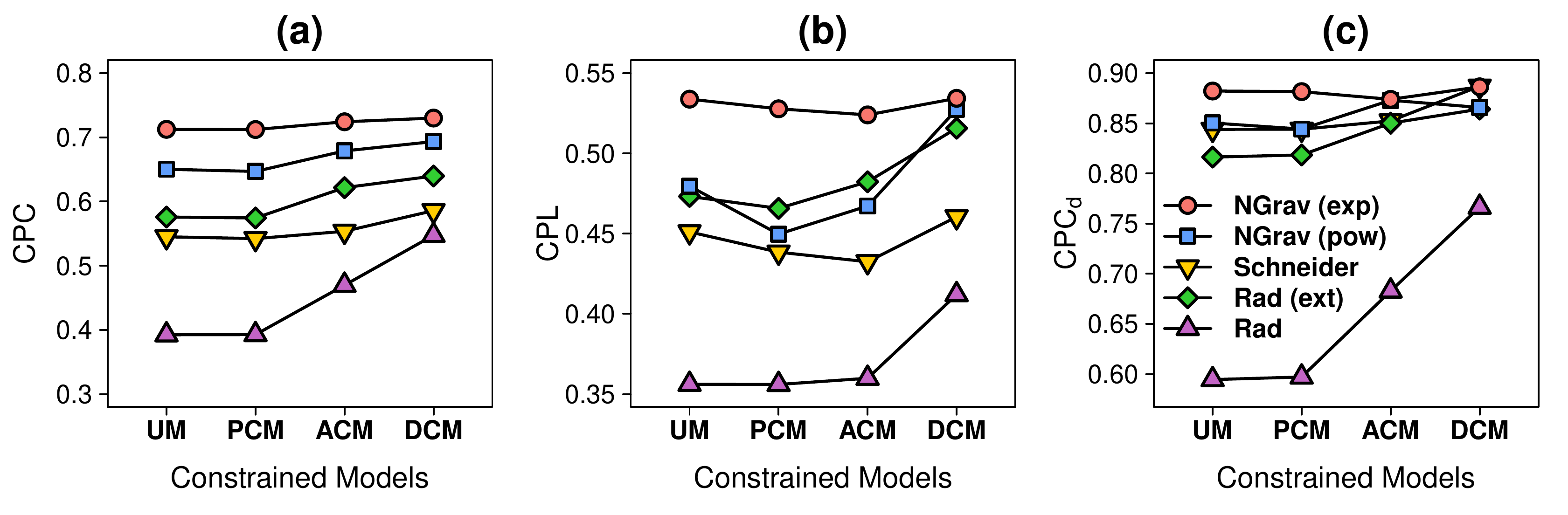}
\caption{\textbf{Performance of the constrained models according to the gravity and the intervening opportunities laws.} (a) Average CPC. (b) Average CPL. (c) Average CPC$_d$. The red circles represent the normalized gravity law with the exponential distance decay function; The blue squares represent the normalized gravity law with the power distance decay function; The point down triangles represent the Schneider's intervening opportunities law; The green diamonds represent the extended radiation law; The purple triangles represent the original radiation law. \label{Fig7}}
\end{center}
\end{figure*}

\subsection*{Structure of the commuting network}

We consider next the capacity of the gravity and the intervening opportunities laws to recover the structure of the empirical commuting networks. Figure \ref{Fig4}b shows the average common part of links obtained with the different laws and models. We observe that the gravity law with an exponential distance decay function outperforms the other laws when the unconstrained and the single constrained models are used to generate the flows. However, when the doubly constrained models is considered, very similar results are obtained except for the Schneider law and the original version of the radiation law. In any case, the common part of links never exceed $0.55$, this can be explained by the fact that, globally, the different laws fail at reproducing the number of links. Indeed, as it can be seen in Figure \ref{Fig5}, which displays the ratio between the number of links generated with the models and the observed ones, the radiation law and the exponential gravity law tend to underestimate the number of links whereas the extended radiation law and the power gravity law overestimate it. The flows networks generated with the Schneider law have globally a number of links closer to the observed values than the networks generated with the other laws.

\subsection*{Commuting distance distribution} 

Another important feature to study is the commuting distance distribution. Figure \ref{Fig4}c shows the average common part of commuters according to the distance obtained with the different models and laws. The results obtained with the exponential gravity law are slightly better than the ones obtained with the other laws. However, the results are globally good, and except the original radiation law, the gravity and intervening opportunities laws are able to reproduce more than $80\%$ of the commuting distances.

To go further, we plot in Figure \ref{Fig6} the observed and the simulated commuting distance distributions obtained with the production constrained model in France and United States. We can clearly see that the exponential gravity law is better for estimating commuting distances which are below a certain threshold equal to $50$ km in France and $150$ km in United States. After this threshold, it fails at estimating the commuting flows as it is the case for the Schneider's intervening opportunities law. On the contrary, the radiation laws and the gravity law with a power distance decay function are able to estimate commuting flows at large distances. However, we have to keep in mind that the proportion of commuters traveling such long distances are less than $6\%$ in France and $5\%$ in United States. Besides, one can legitimately wonder whether these long travels are repeated twice per day or if they may be an artifact of the way in which the census information is collected.

\subsection*{Robustness against changes in the inputs}

In Equations \ref{grav} and \ref{IO}, the population is used as input instead of the outflows $O_i$ and the inflows $D_j$, which are usually preferred since they are a more faithful reflection of the job demand and offer. The job demand and offer are considered to be related to the population but the proportion is rarely direct (it needs to be adjusted with an exponent) and according to the case study, the fit can be bad. In order to assess the robustness of the results to changes in the input data, we consider the results obtained with the gravity law (Equation \ref{gravOi}) and the general intervening opportunities law (Equation \ref{IOOi}) based on the in and out flows. In the case of the intervening opportunities laws, $s_{ij}$ is the number of in-commuters in a circle of radius $d_{ij}$ centered in $i$ (excluding the source and destination) and the role of the populations in the gravity law is taken by $O_i$ and $D_j$. To be more specific, the gravity law becomes:

\begin{figure*}
\begin{center}
\includegraphics[scale=0.6]{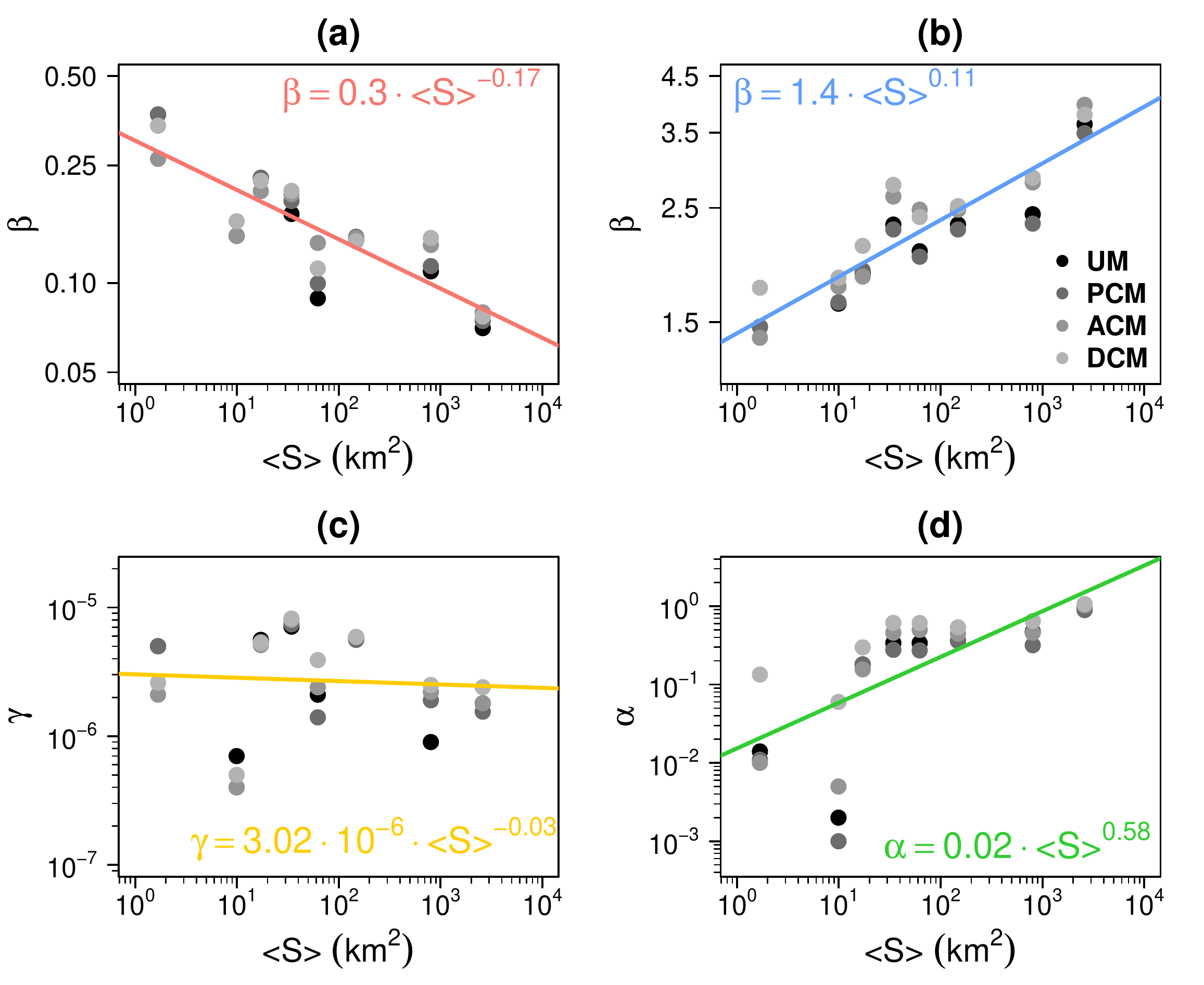}
\caption{\textbf{Parameter value as a function of the average unit surface.} (a) Normalized gravity laws with an exponential distance decay function. (b) Normalized gravity laws with a power distance decay function. (c) Schneider's intervening opportunities law. (d) Extended radiation law.  \label{Fig8}}
\end{center}
\end{figure*} 

\begin{equation}
   p_{ij} \propto O_i \frac{D_j f(d_{ij})}{\sum_{k=1}^n D_k f(d_{ik})},\,\,\,\,\,\,i\ne j \label{gravOi}
\end{equation}

while the intervening opportunities law can be written as

\begin{equation}
   p_{ij} \propto O_i \frac{\mathbb{P}(1|D_i,D_j,s_{ij})}{\sum_{k=1}^n\mathbb{P}(1|D_i,D_k,s_{ik})},\,\,\,\,\,\,i\ne j \label{IOOi}
\end{equation}

Figure \ref{Fig7} displays the average CPC, CPL and CPC$_d$ obtained with the four models according to the laws averaged over the eight case studies. As it can be seen on these plots the results observed in Figure \ref{Fig4} are quite stable to changes in the input data. 

\begin{figure*}
  \begin{center}
		\includegraphics[width=12cm]{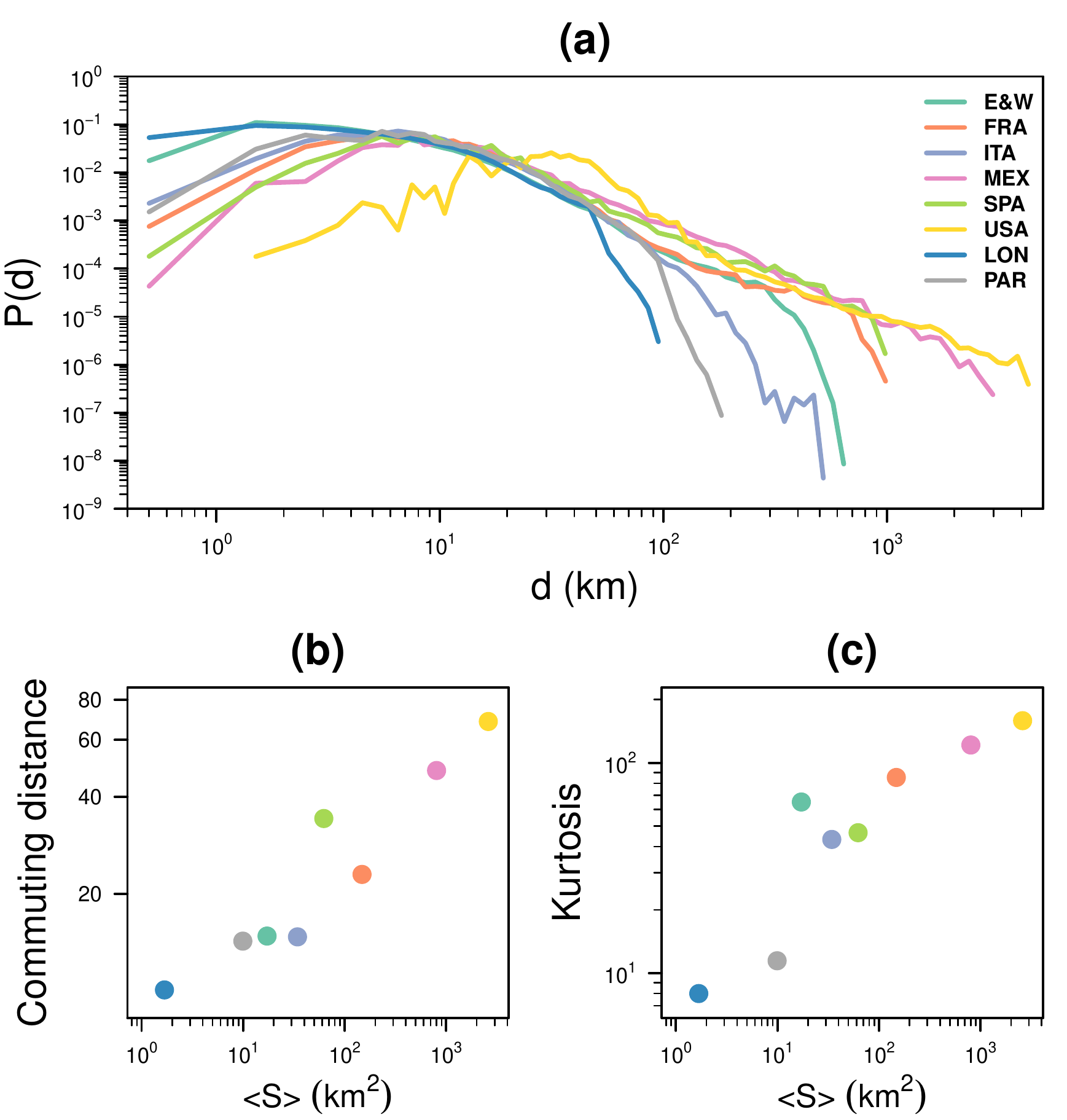}
	\end{center}
	\caption{\textbf{Observed commuting distance distributions.} (a) Probability density function of the commuting distance distribution according to the case study. (b) Average commuting distance as a function of the average unit surface. (c) Pearson's measure of Kurtosis as a function of the average unit surface. \label{Fig9}}
\end{figure*}

\subsection*{Parameter calibration in the absence of detailed data}

An important issue with the estimation of commuting flows is the calibration of the parameters. Indeed, how to calibrate the parameters $\beta$, $\gamma$ and $\alpha$ in the absence of detailed data? This problem has already been tackled in previous studies \cite{Balcan2009,Lenormand2012,Yang2014}. In \cite{Lenormand2012}, the authors have shown that, in the case of the exponential form of the gravity law, the value of $\beta$ can be directly inferred from the average census unit surface with the relationship $\beta=0.3\,<S>^{-0.18}$. Similarly, \cite{Yang2014} proposed to estimate the value of $\alpha$ in the extended radiation law with the average spatial scale $l=\sqrt{<S>}$ using the functional relationship $\alpha=0.0085\,l^{1.33}$.

In Figure \ref{Fig8}, we plot the calibrated value of $\beta$, $\gamma$ and $\alpha$ obtained with the laws based on the population as a function of the average census unit surface $<S>$ for the four constrained models. Figure \ref{Fig8}a shows the relationship obtained with the gravity law with an exponential distance decay function. We observe that the coefficients of the relationship are the same than the ones obtained in \cite{Lenormand2012}. This is not surprising since three datasets out of the six used here coincide. In this case, the value of $\beta$ decreases with larger spatial scales. This can be explained by the fact that $\beta$ in the exponential form of the gravity law is inversely proportional to the average commuting distance and such distance increases with the average unit surface since the shorter distance trips are excluded (Figure \ref{Fig9}b). Figure \ref{Fig8}b displays the same relationship for the power form of the gravity law, in this case the value of $\beta$ increases with the scale to fit the tail of the commuting distance distribution. In fact, we observe in the data that, globally, the steepness of the curve (measured with the Pearson's Kurtosis) increases with the scale (Figure \ref{Fig9}c). Figure \ref{Fig8}c shows the results obtained with the parameter $\gamma$ of the Schneider intervening opportunities law. The value of $\gamma$ seems to decrease slightly with the scale but the existence of a relationship between the two variables is not significant. Finally, we plot in Figure \ref{Fig8}d the relationship between the parameter $\alpha$ of the extended radiation law and the average unit surface, the exponent obtained is similar to the one reported in \cite{Yang2014}. In the extended version of the radiation law, the parameter $\alpha$ controls the effect of the number of job opportunities between home and work on the job selection. In particular, for a given number of job opportunities, higher the value of $\alpha$, higher the probability of not accepting a job among these opportunities. This implies that $\alpha$ is directly proportional to the average commuting distance and, by extension, to the average unit surface (Figure \ref{Fig9}b). As mentioned in \cite{Yang2014}, the value of $\alpha$ is also influenced by the heterogeneity of the distribution of opportunities. As it can be seen in Figure \ref{Fig8}d, the three case studies presenting the largest deviation from the regression line are also the most heterogeneous ones (Paris, Spain and Italy which have the second, fourth and fifth smallest average unit surface, respectively).

\begin{figure}
  \begin{center}
		\includegraphics[width=\linewidth]{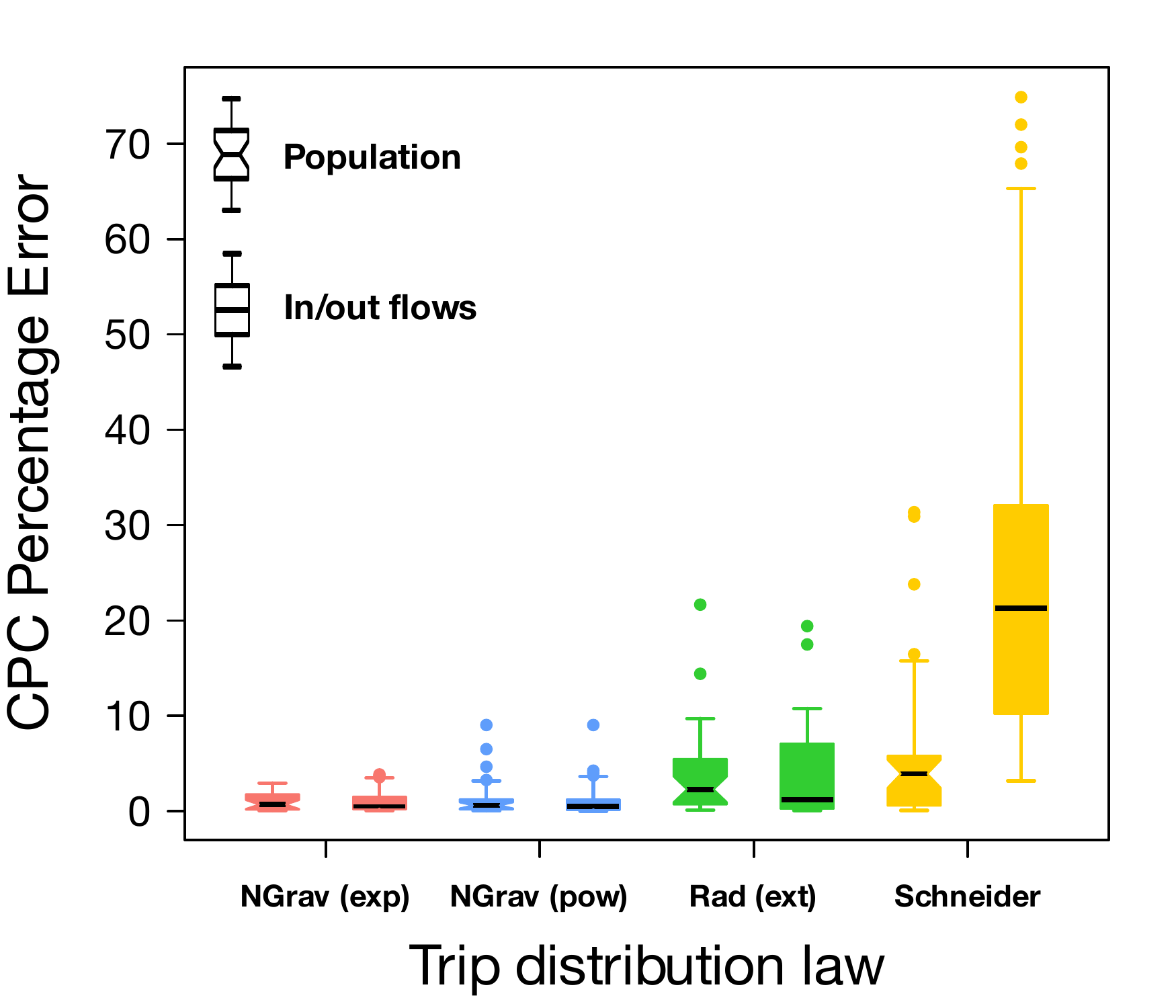}
	\end{center}
	\caption{\textbf{CPC absolute percentage error.} Boxplots of the absolute percentage error between the CPC obtained with a calibrated value of the parameters and the CPC obtained with values estimated with the regression models obtained with the laws based on the population. The notched and classic boxplots represent the percentage error obtained with the laws based on the population and the number of in/out flows, respectively. The boxplot is composed of the first decile, the lower hinge, the median, the upper hinge and the last decile. \label{Fig10}}
\end{figure}

As in \cite{Lenormand2012}, it is possible to assess the quality of the parameter estimation by measuring its impact on the CPC. The idea is to measure for each law, model and case study, the difference between the CPC obtained with the calibrated value of the parameter and the CPC obtained with the estimated one. The parameter value is estimated with the regression model obtained with the laws based on the population and the difference between the original CPC and the ``estimated'' one is measured with the absolute percentage error (i.e. absolute error as a percent of the original CPC value). In order to assess the robustness of the estimation in changes in the input we have also measured the CPC percentage error obtained with an estimation of the parameters for the laws based on the in/out flows. Note that in this case the parameters' estimation come also from regression models obtained with the laws based on the population. The results are presented in Figure \ref{Fig10}. The CPC percentage errors obtained with the gravity laws are globally small and robust to the change of inputs. They vary at most by $4\%$ of the original CPC values for the exponential form and $10\%$ for the power form. Similar results are obtained for the extended radiation law where the majority of the errors are below $10\%$ and vary at most by $22\%$ of the original CPC values. This means that for these laws the parameter value can be directly inferred from the scale, and thus, commuting networks at different scales can be generated without requiring detailed data for calibration. The situation is different for the Schneider's intervening opportunities law very sensible to change in inputs. For the law based on the population, the errors obtained for the CPC are reasonable, the majority of them are below $10\%$. However when we try to estimate the value of $\gamma$ for the law based on in/out flows with a regression model obtained with the law based on the population the CPC percentage error increases dramatically, meaning that the value of $\gamma$ is highly dependent on the variable uses as a surrogate measure of the number of ``real'' opportunities. 

\section*{DISCUSSION}

In summary, we have compared different versions of the gravity and the intervening opportunities laws. These two approaches have already been compared in the past but using different inputs, number of parameters and/or type of constraints. For this reason, the aim of this work has been to bring some light into the discussion by systematically comparing the intervening opportunities and the gravity laws taking care of dissociating the probabilistic laws and the constrained models used to generate the trip networks. We have shown that, globally, the gravity approach outperforms the intervening opportunities approach to estimate the commuting flows but also to preserve the commuting network structure and to fit of the commuting distance distribution. More particularly the gravity law with the exponential distance decay function give better results than the other laws even if it fails at estimating commuting flows at large distances. The reason for this is that most of the travels are short-range, which are better capture by the gravity law with exponential decay in the distance. The large distance commuting trips are few and probably associated with weekly rather  than daily commuting. To handle these different types of mobility, it may be necessary to investigate further the nature of the trips and to consider even mixed models for different displacement lengths. The superiority of the gravity law is very robust to the choice of goodness-of-fit measure and to the change of input. Regarding a more practical issue which is the calibration of the parameters without detailed data, we shown that the parameter values can be estimated with the average unit surface. We also demonstrated that, except for the Schneider's intervening opportunities law, this estimation is robust to changes in input data. This allows for a direct estimation of the commuting flows even in the absence of detailed data for calibration.

Although more research is needed to investigate the link between mobility, distances and intervening opportunities for other types of movements such as migrations, tourism or freight distribution, the distance seems to play a more important role than the number of intervening opportunities in work location choices. More specifically, the superiority of the gravity approach seems to be due to its flexibility, and, what was considered as a weakness by \cite{Simini2012}, the lack of theoretical guidance to choose the distance-decay function, emerges as a strength. Indeed, people do not choose their place of work as they choose their new place of residence, therefore, having the possibility of adjusting the effect of the distance in the decision process is clearly an advantage which does not apply to the intervening opportunities approach in its present form. 

The objective of this work has been to establish the basis for a fair and systematic comparison separating probabilistic laws and different degrees of constraint trip generation models. Our results emphasize the importance of identifying and separating the different processes involved in the estimation of flows between locations for the comparison of spatial interaction models. Indeed, the use of these models in contexts such as urban and infrastructure planning, where large investments are at stake, imposes the need for the selection of the aptest model before taking decisions based on its results. The software package to generate spatial networks using the approach described in the paper can be downloaded from \url{https://github.com/maximelenormand/Trip-distribution-laws-and-models}.

\section*{ACKNOWLEDGEMENTS}

Partial financial support has been received from the Spanish Ministry of Economy (MINECO) and FEDER (EU) under the project INTENSE@COSYP (FIS2012-30634),  and from the EU Commission through projects INSIGHT. The work of ML has been funded under the PD/004/2013 project, from the Conselleria de Educación, Cultura y Universidades of the Government of the Balearic Islands and from the European Social Fund through the Balearic Islands ESF operational program for 2013-2017. JJR acknowledges funding from the Ram\'on y Cajal program of MINECO.
 
\bibliographystyle{unsrt}
\bibliography{RadGrav}

\begin{table*}[t]
  \fontsize{6}{6}\selectfont
  \caption{\label{tab2}\textbf{Law exhibiting the best performances according to the inputs, case studies, constrained models and goodness-of-fit measures.}}
  \begin{center}
    \begin{tabular}{llllllll}
		  \hline 
      \textbf{Inputs} & \textbf{Case study} & \textbf{Model} & \textbf{CPC} & \textbf{CPL} & \textbf{CPC$_d$} & \textbf{NRMSE} & \textbf{I}\\
\hline 
Population	&	E\&W	&	UM	&	NGrav (exp)	&	NGrav (exp)	&	IO	&	NGrav (exp)	&	IO	\\
Population	&	FRA	&	UM	&	NGrav (exp)	&	NGrav (exp)	&	NGrav (exp)	&	Rad (ext)	&	NGrav (exp)	\\
Population	&	ITA	&	UM	&	NGrav (exp)	&	NGrav (exp)	&	IO	&	Rad (ext)	&	NGrav (exp)	\\
Population	&	MEX	&	UM	&	NGrav (pow)	&	NGrav (exp)	&	Rad	&	Rad (ext)	&	NGrav (exp)	\\
Population	&	SPA	&	UM	&	NGrav (pow)	&	NGrav (exp)	&	NGrav (pow)	&	Rad (ext)	&	NGrav (exp)	\\
Population	&	USA	&	UM	&	NGrav (exp)	&	NGrav (exp)	&	NGrav (pow)	&	Rad (ext)	&	NGrav (exp)	\\
Population	&	LON	&	UM	&	NGrav (exp)	&	IO	&	NGrav (pow)	&	NGrav (exp)	&	NGrav (exp)	\\
Population	&	PAR	&	UM	&	NGrav (exp)	&	NGrav (exp)	&	NGrav (pow)	&	NGrav (pow)	&	NGrav (exp)	\\
\hline
Population	&	E\&W	&	PCM	&	NGrav (exp)	&	NGrav (exp)	&	IO	&	NGrav (exp)	&	IO	\\
Population	&	FRA	&	PCM	&	NGrav (exp)	&	NGrav (exp)	&	NGrav (exp)	&	NGrav (exp)	&	NGrav (exp)	\\
Population	&	ITA	&	PCM	&	NGrav (exp)	&	NGrav (exp)	&	IO	&	NGrav (exp)	&	NGrav (exp)	\\
Population	&	MEX	&	PCM	&	NGrav (exp)	&	NGrav (exp)	&	Rad (ext)	&	NGrav (exp)	&	NGrav (exp)	\\
Population	&	SPA	&	PCM	&	NGrav (exp)	&	NGrav (exp)	&	NGrav (pow)	&	NGrav (exp)	&	NGrav (exp)	\\
Population	&	USA	&	PCM	&	NGrav (exp)	&	NGrav (exp)	&	NGrav (exp)	&	NGrav (exp)	&	NGrav (exp)	\\
Population	&	LON	&	PCM	&	NGrav (exp)	&	IO	&	NGrav (pow)	&	NGrav (exp)	&	NGrav (exp)	\\
Population	&	PAR	&	PCM	&	NGrav (exp)	&	NGrav (exp)	&	NGrav (exp)	&	NGrav (exp)	&	NGrav (exp)	\\
Population	&	E\&W	&	ACM	&	NGrav (exp)	&	NGrav (exp)	&	IO	&	NGrav (exp)	&	NGrav (exp)	\\
\hline
Population	&	FRA	&	ACM	&	NGrav (exp)	&	Rad (ext)	&	NGrav (exp)	&	NGrav (exp)	&	NGrav (exp)	\\
Population	&	ITA	&	ACM	&	NGrav (exp)	&	NGrav (exp)	&	IO	&	NGrav (pow)	&	NGrav (exp)	\\
Population	&	MEX	&	ACM	&	NGrav (exp)	&	NGrav (exp)	&	NGrav (exp)	&	Rad (ext)	&	NGrav (exp)	\\
Population	&	SPA	&	ACM	&	NGrav (exp)	&	NGrav (pow)	&	NGrav (pow)	&	NGrav (exp)	&	NGrav (exp)	\\
Population	&	USA	&	ACM	&	NGrav (exp)	&	NGrav (pow)	&	NGrav (exp)	&	NGrav (exp)	&	NGrav (exp)	\\
Population	&	LON	&	ACM	&	NGrav (exp)	&	NGrav (exp)	&	NGrav (pow)	&	NGrav (exp)	&	NGrav (exp)	\\
Population	&	PAR	&	ACM	&	NGrav (exp)	&	NGrav (exp)	&	NGrav (pow)	&	IO	&	NGrav (exp)	\\
\hline
Population	&	E\&W	&	DCM	&	NGrav (exp)	&	NGrav (exp)	&	IO	&	NGrav (exp)	&	NGrav (exp)	\\
Population	&	FRA	&	DCM	&	NGrav (exp)	&	Rad (ext)	&	NGrav (exp)	&	NGrav (exp)	&	NGrav (exp)	\\
Population	&	ITA	&	DCM	&	NGrav (exp)	&	NGrav (exp)	&	NGrav (pow)	&	NGrav (exp)	&	NGrav (exp)	\\
Population	&	MEX	&	DCM	&	NGrav (exp)	&	NGrav (pow)	&	Rad (ext)	&	NGrav (exp)	&	NGrav (exp)	\\
Population	&	SPA	&	DCM	&	NGrav (pow)	&	NGrav (pow)	&	NGrav (pow)	&	NGrav (exp)	&	NGrav (exp)	\\
Population	&	USA	&	DCM	&	NGrav (exp)	&	Rad (ext)	&	NGrav (exp)	&	NGrav (exp)	&	NGrav (exp)	\\
Population	&	LON	&	DCM	&	NGrav (exp)	&	NGrav (exp)	&	NGrav (exp)	&	NGrav (exp)	&	NGrav (exp)	\\
Population	&	PAR	&	DCM	&	NGrav (exp)	&	NGrav (exp)	&	NGrav (pow)	&	NGrav (exp)	&	NGrav (exp)	\\
\hline
In/out flows	&	E\&W	&	UM	&	NGrav (exp)	&	NGrav (exp)	&	IO	&	NGrav (exp)	&	NGrav (exp)	\\
In/out flows	&	FRA	&	UM	&	NGrav (exp)	&	NGrav (pow)	&	NGrav (exp)	&	NGrav (exp)	&	NGrav (exp)	\\
In/out flows	&	ITA	&	UM	&	NGrav (exp)	&	NGrav (exp)	&	IO	&	NGrav (exp)	&	NGrav (exp)	\\
In/out flows	&	MEX	&	UM	&	NGrav (exp)	&	NGrav (exp)	&	Rad (ext)	&	NGrav (exp)	&	NGrav (exp)	\\
In/out flows	&	SPA	&	UM	&	NGrav (exp)	&	NGrav (exp)	&	NGrav (pow)	&	NGrav (exp)	&	NGrav (exp)	\\
In/out flows	&	USA	&	UM	&	NGrav (exp)	&	NGrav (exp)	&	NGrav (exp)	&	NGrav (exp)	&	NGrav (exp)	\\
In/out flows	&	LON	&	UM	&	NGrav (exp)	&	NGrav (exp)	&	NGrav (pow)	&	NGrav (exp)	&	NGrav (exp)	\\
In/out flows	&	PAR	&	UM	&	NGrav (exp)	&	NGrav (exp)	&	NGrav (pow)	&	NGrav (exp)	&	NGrav (exp)	\\
\hline
In/out flows	&	E\&W	&	PCM	&	NGrav (exp)	&	NGrav (exp)	&	IO	&	NGrav (exp)	&	NGrav (exp)	\\
In/out flows	&	FRA	&	PCM	&	NGrav (exp)	&	NGrav (exp)	&	NGrav (exp)	&	NGrav (exp)	&	NGrav (exp)	\\
In/out flows	&	ITA	&	PCM	&	NGrav (exp)	&	NGrav (exp)	&	IO	&	NGrav (exp)	&	NGrav (exp)	\\
In/out flows	&	MEX	&	PCM	&	NGrav (exp)	&	NGrav (exp)	&	Rad (ext)	&	NGrav (exp)	&	NGrav (exp)	\\
In/out flows	&	SPA	&	PCM	&	NGrav (exp)	&	NGrav (exp)	&	NGrav (exp)	&	NGrav (exp)	&	NGrav (exp)	\\
In/out flows	&	USA	&	PCM	&	NGrav (exp)	&	NGrav (exp)	&	NGrav (exp)	&	NGrav (exp)	&	NGrav (exp)	\\
In/out flows	&	LON	&	PCM	&	NGrav (exp)	&	NGrav (exp)	&	NGrav (exp)	&	NGrav (exp)	&	NGrav (exp)	\\
In/out flows	&	PAR	&	PCM	&	NGrav (exp)	&	NGrav (exp)	&	NGrav (exp)	&	NGrav (exp)	&	NGrav (exp)	\\
\hline
In/out flows	&	E\&W	&	ACM	&	NGrav (exp)	&	NGrav (exp)	&	IO	&	NGrav (exp)	&	NGrav (exp)	\\
In/out flows	&	FRA	&	ACM	&	NGrav (exp)	&	NGrav (exp)	&	NGrav (exp)	&	NGrav (exp)	&	NGrav (exp)	\\
In/out flows	&	ITA	&	ACM	&	NGrav (exp)	&	NGrav (exp)	&	IO	&	NGrav (pow)	&	NGrav (exp)	\\
In/out flows	&	MEX	&	ACM	&	NGrav (exp)	&	NGrav (exp)	&	NGrav (pow)	&	NGrav (exp)	&	NGrav (exp)	\\
In/out flows	&	SPA	&	ACM	&	NGrav (exp)	&	NGrav (exp)	&	NGrav (pow)	&	NGrav (exp)	&	NGrav (exp)	\\
In/out flows	&	USA	&	ACM	&	NGrav (exp)	&	NGrav (exp)	&	NGrav (exp)	&	NGrav (exp)	&	NGrav (exp)	\\
In/out flows	&	LON	&	ACM	&	NGrav (exp)	&	Rad (ext)	&	IO	&	NGrav (exp)	&	NGrav (exp)	\\
In/out flows	&	PAR	&	ACM	&	NGrav (exp)	&	NGrav (exp)	&	NGrav (pow)	&	NGrav (exp)	&	NGrav (exp)	\\
\hline
In/out flows	&	E\&W	&	DCM	&	NGrav (exp)	&	NGrav (exp)	&	IO	&	NGrav (exp)	&	NGrav (exp)	\\
In/out flows	&	FRA	&	DCM	&	NGrav (exp)	&	NGrav (pow)	&	NGrav (exp)	&	NGrav (exp)	&	NGrav (exp)	\\
In/out flows	&	ITA	&	DCM	&	NGrav (exp)	&	NGrav (exp)	&	IO	&	NGrav (exp)	&	NGrav (exp)	\\
In/out flows	&	MEX	&	DCM	&	NGrav (exp)	&	NGrav (pow)	&	Rad (ext)	&	NGrav (exp)	&	NGrav (exp)	\\
In/out flows	&	SPA	&	DCM	&	NGrav (pow)	&	NGrav (pow)	&	NGrav (pow)	&	NGrav (exp)	&	NGrav (exp)	\\
In/out flows	&	USA	&	DCM	&	NGrav (exp)	&	Rad (ext)	&	NGrav (exp)	&	NGrav (exp)	&	NGrav (exp)	\\
In/out flows	&	LON	&	DCM	&	NGrav (exp)	&	NGrav (exp)	&	NGrav (exp)	&	NGrav (exp)	&	NGrav (exp)	\\
In/out flows	&	PAR	&	DCM	&	NGrav (exp)	&	NGrav (exp)	&	NGrav (pow)	&	NGrav (exp)	&	NGrav (exp)	\\																
\hline
    \end{tabular}
  \end{center}
\end{table*}

\end{document}